\documentclass[preprint,aps,tightenlines,showpacs,superscriptaddress]{revtex4}
\usepackage[dvips]{graphicx}
\usepackage{dcolumn}
\usepackage{bm}
\usepackage{epsfig}

\newcommand{\beq}{\begin{equation}}
\newcommand{\eeq}{\end{equation}}

\begin{document}
\title{Dynamical Coupled-Channels Effects on Pion Photoproduction 
\footnote{Notice: Authored by Jefferson Science Associates, LLC under U.S. DOE 
Contract No. DE-AC05-06OR23177. The U.S. Government retains a non-exclusive,
paid-up, irrevocable, world-wide license to publish or reproduce this
manuscript for U.S. Government purposes. }} 

\vspace{0.5cm}
\author{B. Juli\'a-D\'{\i}az} 
\affiliation{ Excited Baryon Analysis Center (EBAC), Thomas Jefferson National
Accelerator Facility, Newport News, VA 22901, USA}
\affiliation{Departament d'Estructura i Constituents de la Mat\`{e}ria,
Universitat de Barcelona, E--08028 Barcelona, Spain}
\author{T.-S. H. Lee}
\affiliation{ Excited Baryon Analysis Center (EBAC), Thomas Jefferson National
Accelerator Facility, Newport News, VA 22901, USA}
\affiliation{Physics Division, Argonne National Laboratory, 
Argonne, IL 60439, USA}
\author{A. Matsuyama}
\affiliation{ Excited Baryon Analysis Center (EBAC), Thomas Jefferson National
Accelerator Facility, Newport News, VA 22901, USA}

\affiliation{Department of Physics, Shizuoka University, Shizuoka 422-8529, Japan}
\author{T. Sato}
\affiliation{ Excited Baryon Analysis Center (EBAC), Thomas Jefferson National
Accelerator Facility, Newport News, VA 22901, USA}
\affiliation{Department of Physics, Osaka University, Toyonaka, 
Osaka 560-0043, Japan}

\author{L.C. Smith}
\affiliation{ Excited Baryon Analysis Center (EBAC), Thomas Jefferson National
Accelerator Facility, Newport News, VA 22901, USA}
\affiliation{Department of Physics,  University of Virginia, VA 22901,
USA}

\begin{abstract}
The electromagnetic pion production reactions are investigated 
within the dynamical coupled-channels model developed in 
{\bf Physics Reports, 439, 193 (2007)}.
The meson-baryon channels included in this study are $\gamma N$, $\pi N$, 
$\eta N$, and the $\pi\Delta$, $\rho N$ and $\sigma N$ resonant 
components of the $\pi\pi N$ channel. With the hadronic 
parameters of the model determined in a recent study of $\pi N$ 
scattering, we show that the pion photoproduction data up to the 
second resonance region can be described to a very large extent 
by only adjusting the bare $\gamma N \rightarrow N^*$ helicity 
amplitudes, while the non-resonant electromagnetic couplings are 
taken from previous works. It is found that the coupled-channels
effects can contribute about 30 - 40 $\%$ of the production cross 
sections in the $\Delta$ (1232) resonance region, and can drastically 
change the magnitude and shape of the cross sections in the 
second resonance region. The importance of the off-shell effects 
in a dynamical approach is also demonstrated. The meson cloud 
effects as well as the coupled-channels contributions to the 
$\gamma N \rightarrow N^*$ form factors are found to be mainly 
in the low $Q^2$ region. For the magnetic M1 
$\gamma N \rightarrow \Delta$ (1232) form factor, the results are 
close to that of the Sato-Lee Model. Necessary improvements to 
the model and future developments are discussed.

\end{abstract}
                                                                                
\maketitle

\section{Introduction}
It is well recognized~\cite{lee-reviewa,lee-reviewb} that the data 
of electromagnetic meson production reactions can be used to reveal 
the structure of the excited states ($N^*$) of the nucleon. In this 
paper, we report on an investigation in this direction within the 
dynamical coupled-channels model (MSL) presented in Ref.~\cite{msl} 
which is being applied at the Excited Baryon Analysis Center (EBAC) 
of Jefferson Laboratory.
 
The coupled-channels approach has been used~\cite{giessen,kvi,bonn,yang1} 
in recent years to analyze the meson production reaction data. It 
is therefore useful to briefly emphasize here the essence of taking 
a dynamical approach as developed in Ref.~\cite{msl} and in earlier 
works~\cite{yang,nbl,gross,sl,ky,julich,afnan,fuda,pasc,cstl,jslt,jlss}.
Since $N^*$ states are unstable, their structure must couple with the 
reaction channels in the meson production reactions. To determine 
correctly the spectrum of $N^*$ states, an analysis of the meson 
production data must account for the coupled-channels unitary condition. 
The extracted $N^*$ parameters can be interpreted correctly only when 
the reaction mechanisms in the short-range region, where we want to map 
out the $N^*$ structure, have been accounted for. The MSL model meets 
these two crucial requirements and is therefore suitable for analyzing 
the world data of meson production reactions induced by pions, photons, 
and electrons.

\begin{figure}[b]
\includegraphics[width=14cm]{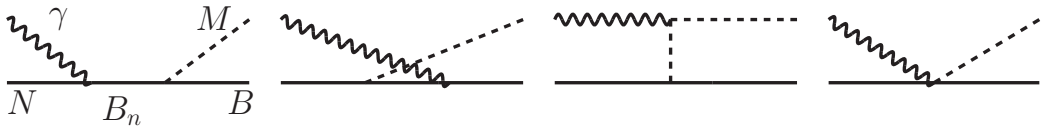}
\caption{The non-resonant electromagnetic meson production interaction
 $v_{MB,\gamma N}$, where $MB =  \pi N, \eta N, \pi\Delta, \rho N, \sigma N$. 
The details are given in Ref.~\cite{msl}.\label{fig:mech}}
\end{figure}

Schematically, the MSL model solves the following coupled integral 
equations in each partial wave
\begin{eqnarray}
T_{\alpha,\beta}(p_\alpha,p_\beta;E)= V_{\alpha,\beta}(p_\alpha,p_\beta) 
+ \sum_{\delta}
 \int p^{2}d p  V_{\alpha,\delta}(p_\alpha, p )
G_{\delta}( p ,E)
T_{\delta,\beta}( p  ,p_\beta,E)  \,,
\label{eq:teq} 
\end{eqnarray}
with
\begin{eqnarray}
V_{\alpha,\beta}(p_\alpha,p_\beta)= v_{\alpha,\beta}(p_\alpha,p_\beta)+
\sum_{N^*}\frac{\Gamma^{\dagger}_{N^*,\alpha}(p_\alpha)
 \Gamma_{N^*,\beta}(p_\beta)} {E-M^0_{N^*}} \,,
\label{eq:veq}
\end{eqnarray}
where $\alpha,\beta,\delta = \gamma N, \pi N, \eta N,$ and 
$\pi\pi N$ which has $\pi \Delta, \rho N, \sigma N$ resonant components, 
$G_\delta (p,E)$ is the propagator of channel $\delta$, $M^0_{N^*}$ 
is the mass of a bare excited nucleon state $N^*$, $v_{\alpha,\beta}$ 
is defined by meson-exchange mechanisms, and the $N^*\rightarrow \beta$
 vertex interaction
$\Gamma_{N^*,\beta}$ is 
related to the quark-gluon sub-structure of $N^*$. If we take the 
on-shell approximation, Eq.~(\ref{eq:teq}) is reduced to the following 
algebraic form of K-matrix models~\cite{said,maid,jlab-yeve,giessen,kvi,bonn}
\begin{eqnarray}
T^{k}_{\alpha,\beta}(p_\alpha,p_\beta,E) &=& 
\sum_{\delta}
V_{\alpha,\delta}(p_\alpha,p_\delta) 
\times[\delta_{\delta, \beta} + i \rho(p_\delta)
T^{k}_{\delta,\beta}(p_\delta,p_\beta,E)] \,,
\label{eq:k-matrix}
\end{eqnarray}
where $\rho(p_\delta)$ is an appropriate phase space factor.
Qualitatively speaking, models that make use of on-shell expressions 
such as Eq.~(\ref{eq:k-matrix}) are used to avoid an explicit 
treatment of the reaction mechanisms in the short range region 
where we want to map out the quark-gluon sub-structure of $N^*$ 
states. Thus the $N^*$ parameters extracted by using Eq.~(\ref{eq:teq}) 
can be more directly interpreted in terms of the quark-gluon 
sub-structure of $N^*$. From the study~\cite{sl,jlss} in the 
$\Delta$ (1232) region, it is reasonable to interpret 
$\Gamma_{N^*,\beta}$ in terms of hadron structure calculations 
with effective degrees of freedom, such as the constituent 
quark model~\cite{capstick} and the model~\cite{roberts} based 
on Dyson-Schwinger Equations. In the near future, one hopes 
to relate $\Gamma_{N^*,\beta}$ to Lattice QCD (LQCD) 
calculations~\cite{dinna,richards}. The possibility of making 
contact with the current hadron structure calculations, which 
can be carried out with sufficient accuracy in the forseeable 
future, is the main motivation for performing the analysis using
a technically much more involved dynamical approach.

To analyze the electromagnetic meson production reactions within 
the MSL model, it is necessary to first determine the hadronic 
part of its Hamiltonian. A progress in this direction has been 
made recently in Ref.~\cite{jlms} (JLMS). The main purpose of this 
work is to explore the consequence of the parameters determined by 
JLMS in describing the pion photoproduction reactions. Within the 
MSL formulation, these hadronic parameters should be consistently
used to fix the strong interaction vertices of the non-resonant 
$\gamma N \rightarrow MB$ interaction $v_{MB,\gamma N}$ of 
Eq.~(\ref{eq:veq}). Since the electromagnetic interactions, such 
as the $\gamma NN, \gamma \pi \rho$, $\gamma\pi\omega$ vertices 
of the considered non-resonant mechanisms illustrated in 
Fig.~\ref{fig:mech}, have been determined in the previous works 
as also given in Ref.~\cite{msl}, the challenge here is to 
explore whether the pion photoproduction data can be described 
by only adjusting the bare $\gamma N \rightarrow N^*$ vertex 
$\Gamma_{N^*,\gamma N}$ of Eq.~(\ref{eq:veq}).

To proceed, we first note that in the study by JLMS~\cite{jlms} 
it was found that the fit to $\pi N$ elastic scattering data
is not sufficient for pinning down precisely the hadronic parameters 
associated with the unstable particle channels $\pi\Delta$, $\rho N$, 
and $\sigma N$. These channels have very large effects in the invariant 
mass $W \geq$ about 1.65 GeV region where the two-pion production 
dominates. It is necessary to include the $\pi N \rightarrow \pi\pi N$ 
data in the fit. This very challenging task is still being pursued at 
EBAC. For this reason, we will limit our investigation to the 
$W =$ 1.1 GeV $-$ 1.65 GeV region which covers the energies of the 
low-lying nucleon resonances in the so called first and second 
resonance regions. The resulting model is sufficient for investigating 
the dynamical coupled-channels effects on the $\gamma N\rightarrow \pi N$
cross sections and the  $\gamma N \rightarrow N^*$ transitions. This 
is the main focus of this work.

The details of the employed dynamical coupled-channels model have been
given in Ref.~\cite{msl}. In section II we only recall the formulas
relevant to the pion photoproduction reactions.
The results  are presented in 
section III. Section IV is devoted to 
discussing the
 necessary improvements and future 
developments.

\section{Formulation}

In the helicity-LSJ mixed-representation~\cite{msl} where the initial 
$\gamma N$ state is specified by its helicities $\lambda_\gamma$ and 
$\lambda_N$ and the final $MB$ states by the $(LS)J$ angular momentum 
variables, the reaction amplitude of $\gamma(\vec{q}) + N (-\vec{q}) \rightarrow
\pi(\vec{k}) + N (-\vec{k})$ at invariant mass $W$ can be written within 
the MSL formulation as (suppress the isospin quantum numbers)
\begin{eqnarray}
T^{J}_{LS_N\pi N,\lambda_\gamma\lambda_N}(k,q,W) = 
{t}^{J}_{L S_N \pi N, \lambda_\gamma\lambda_N}(k, q, W)
+t^{R,J}_{LS_N\pi N,\lambda_\gamma\lambda_N}(k, q, W)\,,
\label{eq:pw-t}
\end{eqnarray}
where $S_N=1/2$ is the nucleon spin, and the non-resonant amplitude is
\begin{eqnarray}
{t}^{J}_{L S_N \pi N,\lambda_\gamma\lambda_N}(k,q,E)&=&
{\it v}^{J}_{L S_N \pi N,\lambda_\gamma\lambda_N}(k,q,E)
+\sum_{M'B'}
\sum_{L^{\prime}S^{\prime}}
\int k^{\prime 2}dk^{\prime }
 {t}^{J}_{L S_N \pi N, L' S'M'B'}(k,k^{\prime},E) \nonumber \\
& & \times G_{M'B'}(k',E)
{\it v}^{J}_{L' S' M'B' , \lambda_\gamma\lambda_N}(k',q,E)\,.
\label{eq:pw-nonr}
\end{eqnarray}
In the above equation, the meson-baryon channels included in the sum
are $M'B'= \pi N, \eta N, \pi\Delta, \rho N, \sigma N$. The matrix 
elements ${\it v}^{J}_{L S MB,\lambda_\gamma\lambda_N}(k,q,E)$, which 
describe the $\gamma N \rightarrow M'B'$ transitions, are calculated 
from the tree-diagrams, illustrated in Fig.~\ref{fig:mech}, of a set 
of phenomenological Lagrangians describing the interactions between
$\gamma$, $\pi$, $\eta$, $\rho$, $\omega$, $\sigma$, $N$, and 
$\Delta$(1232) fields. The details are given explictly in Appendix F 
of Ref.~\cite{msl}.  We will use the hadronic parameters determined 
by JLMS ~\cite{jlms} to evaluate the meson-baryon propagators $G_{M'B'}$ 
and the $\pi N \rightarrow MB$ amplitudes 
${t}^{J}_{L S_N \pi N, L' S'M'B'}(k,k^{\prime},E)$, and to also fix the 
hadronic vertices of the $\gamma N \rightarrow MB$ amplitudes 
${\it v}^{J}_{L S MB,\lambda_\gamma\lambda_N}(k,q,E)$. As discussed in 
section I,  all of the electromagnetic vertices, such as 
$\gamma NN$, $\gamma \pi\pi$, $\gamma \pi\omega$, in 
${\it v}^{J}_{L S MB,\lambda_\gamma\lambda_N}(k,q,E)$ are taken from previous 
works, as also specified in Ref.~\cite{msl}. Thus the non-resonant 
amplitude defined by Eq.~(\ref{eq:pw-nonr}) is completely fixed in the 
present investigation. Such a consistent dynamical treatment of strong 
and electromagnetic reaction mechanisms is highly desirable in using the 
meson production reactions to study the $N^*$ structure. 

\begin{figure}[t]
\centering
\includegraphics[width=12cm,angle=-00]{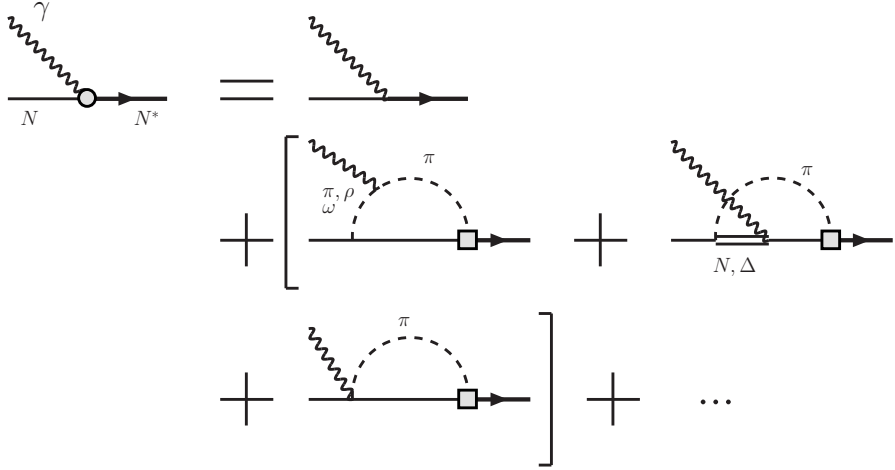}
\caption{Graphical illustration of the contribution to
the $\pi N$ intermediate state to the dressed $\gamma N \rightarrow N^*$ 
vertex defined by Eq.~(\ref{eq:pw-v}).}
\label{fig:d-vertex}
\end{figure}

The resonant amplitude in Eq.~(\ref{eq:pw-t}) is
\begin{eqnarray}
t^{R,J}_{LS_N\pi N,\lambda_\gamma\lambda_N}(k, q, E) =
 \sum_{N^*_i, N^*_j}
[\bar{\Gamma}^{J}_{N^*_i,LS_N\pi N}(k,W)]^*
D_{i,j}(W)
\bar{\Gamma}^{J}_{N^*_j,\lambda_\gamma\lambda_N}(q,W) \,,
\label{eq:pw-r}
\end{eqnarray}
where the dressed vertex functions are defined as
\begin{eqnarray}
[\bar{\Gamma}^{J}_{N^*,LS_N\pi N}(k,W)]^*
&=&[{\Gamma}^{J}_{N^*,LS_N\pi N}(k)]^* \nonumber \\
&+ &\sum_{M'B'}
\sum_{L^{\prime}S^{\prime}}
\int k^{\prime 2}dk^{\prime }
{\it t}^{J}_{L S_N \pi N , L'S' M'B'}(k,k',W)
 G_{M'B'}(k',W)[{\Gamma}^{J}_{N^*,L'S'M'B'}(k')]^* \nonumber \\
\label{eq:pw-pi} \\
\bar{\Gamma}^{J}_{N^*,\lambda_\gamma\lambda_N}(q,W)
&=&{\Gamma}^{J}_{N^*,\lambda_\gamma\lambda_N }(q) \nonumber \\
&+ &\sum_{M'B'}
\sum_{L^{\prime}S^{\prime}}
\int k^{\prime 2}dk^{\prime }
 \bar{\Gamma}^{J}_{N^*,L'S'M'B'}(k',W) G_{M'B'}(k',W)
{\it v}^{J}_{L' S' M'B' , \lambda_\gamma\lambda_N}(k',q)\,.
\label{eq:pw-v}
\end{eqnarray}
The second term of Eq.~(\ref{eq:pw-v}) is due to the mechanism where 
the non-resonant electromagnetic meson production takes place before 
the dressed $N^*$ states are formed. The contribution due to the 
$\pi N$ intermediate state  
is illustrated in Fig.~\ref{fig:d-vertex}. 
Similar to what was defined in Ref.~\cite{sl,jlss}, we call this 
contribution the {\it meson cloud effect} to define precisely what will 
be presented in section III. 

The $N^*$ propagator in Eq.~(\ref{eq:pw-r}) is defined by
\begin{eqnarray}
[D(W)^{-1}]_{i,j} = (W - M^0_{N^*_i})\delta_{i,j} - \bar{\Sigma}_{i,j}(W)\,,
\label{eq:nstar-g}
\end{eqnarray}
where $M_{N^*_i}^0$ is the bare mass of the $N^*_i$ state, 
and the self-energies are
\begin{eqnarray}
\bar{\Sigma}_{i,j}(W)= \sum_{MB}\sum_{LS}\int dk k^2
\Gamma^J_{N_i^*, LS MB}(k) G_{MB}(k,W)
\left[\bar{\Gamma}^J_{N^*_j,LSMB}(k,W)\right]^* \,.
\label{eq:nstar-sigma}
\end{eqnarray}

Since the dressed vertex $\bar{\Gamma}^{J}_{N^*,LS MB}(k,W)$ of 
$N^*\rightarrow MB$ in Eqs.~(\ref{eq:pw-pi})-(\ref{eq:pw-v}) 
and the bare masses $M^0_{N^*}$ and the mass shifts $\bar{\Sigma}_{i,j}$ 
in Eq.~(\ref{eq:nstar-g}) have been determined by JLMS~\cite{jlms}, 
the only unknown quantities in the resonant amplitude 
Eq.~(\ref{eq:pw-r}) are the bare $\gamma N \rightarrow N^*$ vertex 
functions $\Gamma^{J}_{N^*,\lambda_\gamma\lambda_N}(q)$ in Eq.~(\ref{eq:pw-v}). 
We parameterize these functions as  
\begin{eqnarray}
{\Gamma}^{J}_{N^*,\lambda_\gamma\lambda_N}(q)
& =&\frac{1}{(2\pi)^{3/2}}\sqrt{\frac{m_N}{E_N(q)}}\frac{1}{\sqrt{2q}}
[\sqrt{2q_R} A^{J}_{\lambda}]
\delta_{\lambda, (\lambda_\gamma-\lambda_N)} \,,
\label{eq:ggn}
\end{eqnarray}
where $q_R$ is defined by the $N^*$ mass
$M_{N^*} = q_R+E_N(q_R)$. For later discussions, we 
also cast the dressed vertex Eq.~(\ref{eq:pw-v}) into the form of
Eq.~(\ref{eq:ggn}) with a dressed helicity amplitude
\begin{eqnarray}
\bar{A}^{J}_{\lambda}(q) = {A}^{J}_{\lambda} + {A}^{mc,J}_{\lambda}(q)
\label{eq:ggn-d}
\end{eqnarray}
where ${A}^{mc,J}_{\lambda}(q)$ is due to the meson cloud effects
defined by the second term of Eq.~(\ref{eq:pw-v}). The dressed helicity 
amplitude $\bar{A}^{J}_{\lambda}$ is related to the partial width of the 
electromagnetic decay of a $N^*$ resonance defined by
\begin{eqnarray}
[{\rm partial\;decay\;width}]
=\frac{q^2_R}{4\pi}\frac{m_N}{M_{N^*}}\frac{8}{2J+1}
[|\bar{A}^{J}_{3/2}(q_R)|^2+|\bar{A}^{J}_{1/2}(q_R)|^2] \,.
\label{eq:width}
\end{eqnarray}
Eqs.~(\ref{eq:ggn})-(\ref{eq:width}) and Eq.~(\ref{eq:pw-v}) fix the 
relation between the bare helicity amplitude ${A}^{J}_\lambda$ of 
Eq.~(\ref{eq:pw-v}) and the helicity amplitudes $\bar{A}^{J}_\lambda$ 
listed by the Particle Data Group (PDG).

\section{Results}

\begin{figure}
\centering
\includegraphics[width=10cm,angle=-00]{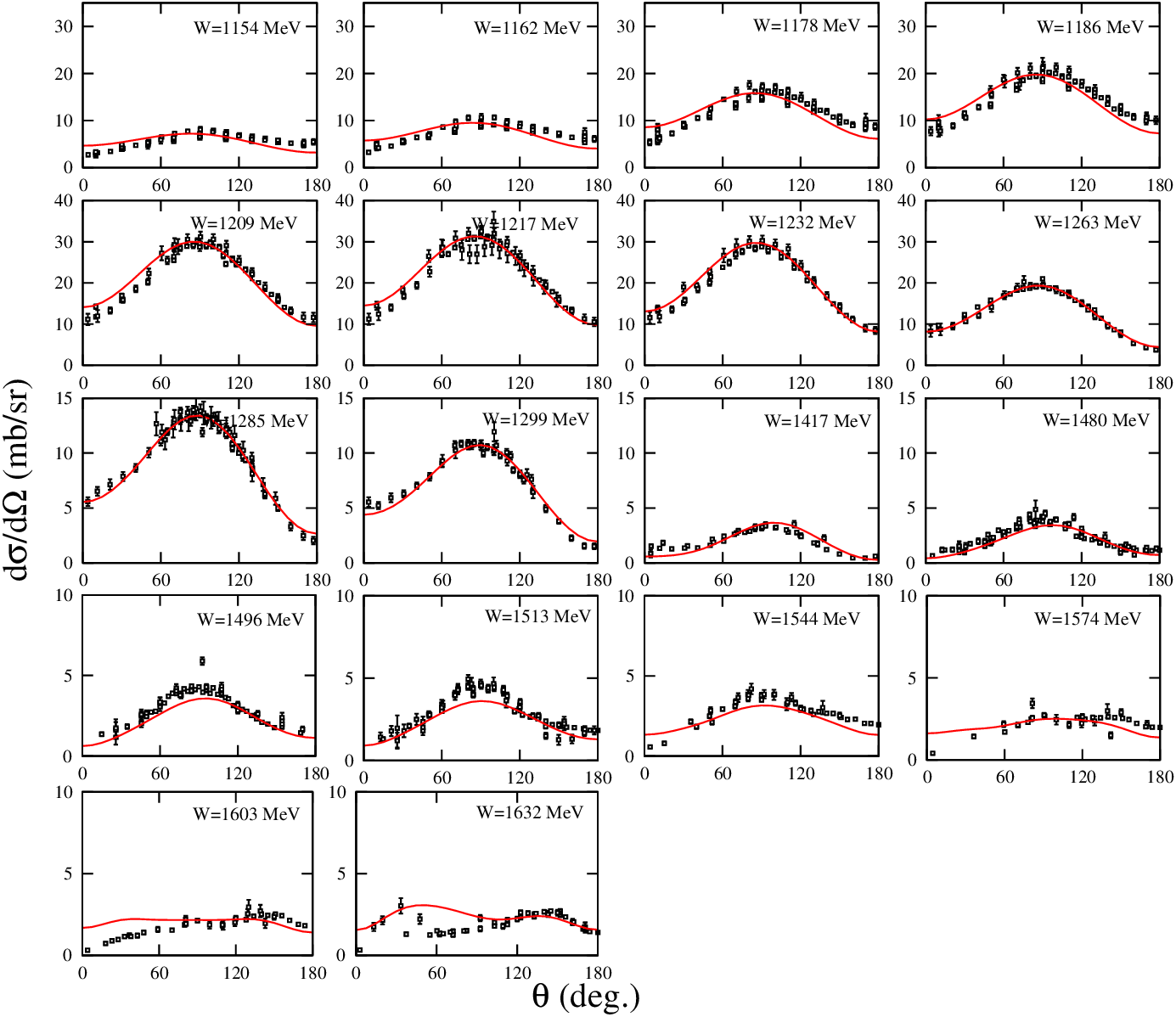}
\caption{Differential cross section for $\gamma p \to \pi^0 p$ compared 
to experimental data obtained from Ref.~\cite{saiddb}.}
\label{fig:dcsp0m0}
\end{figure}

\begin{figure}
\centering
\includegraphics[width=10cm,angle=-00]{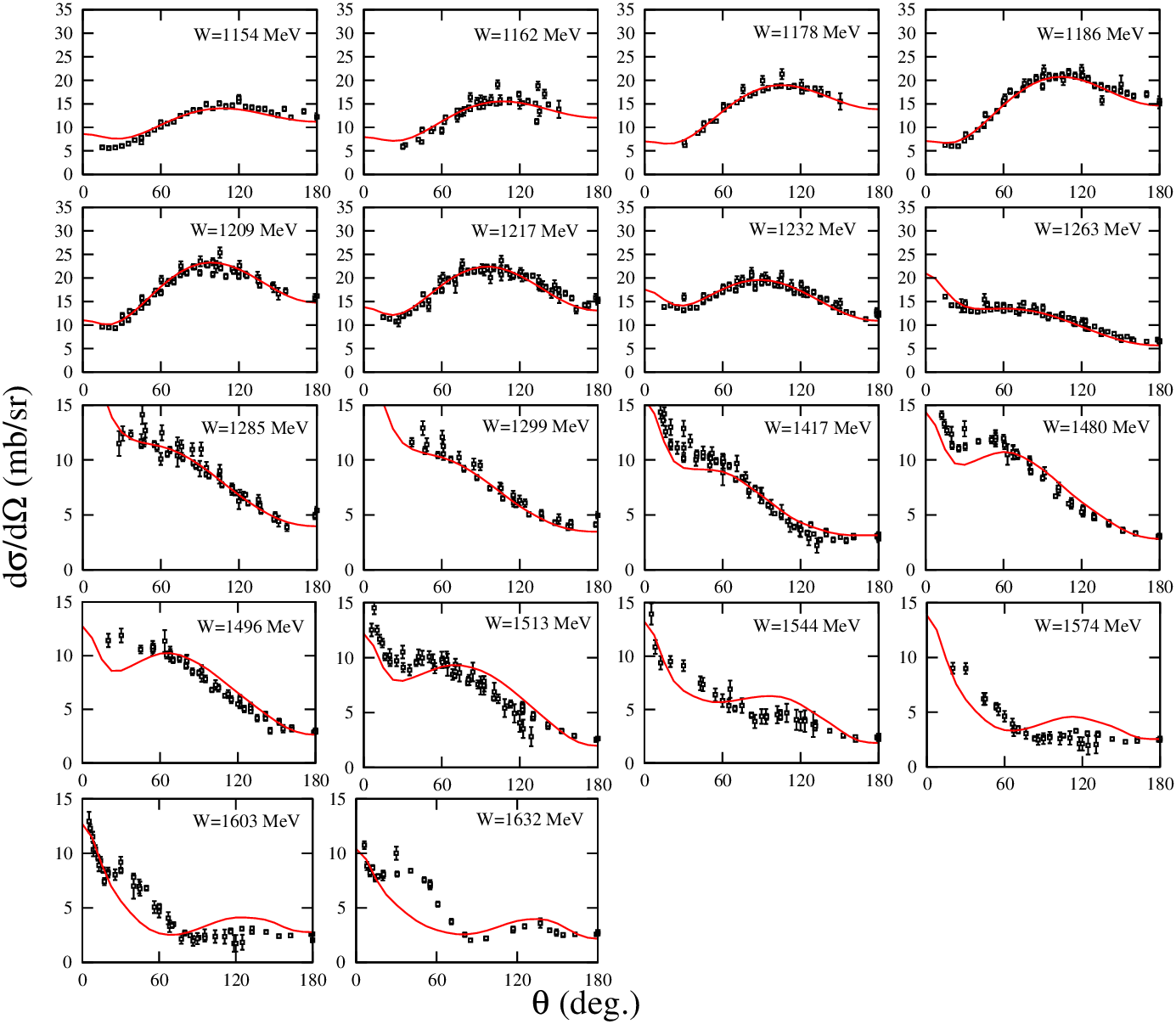}
\caption{Differential cross section for $\gamma p \to \pi^+ n$ compared 
to experimental data obtained from Ref.~\cite{saiddb}.}
\label{fig:dcsppm0}
\end{figure}

\begin{figure}
\centering
\includegraphics[width=10cm,angle=-00]{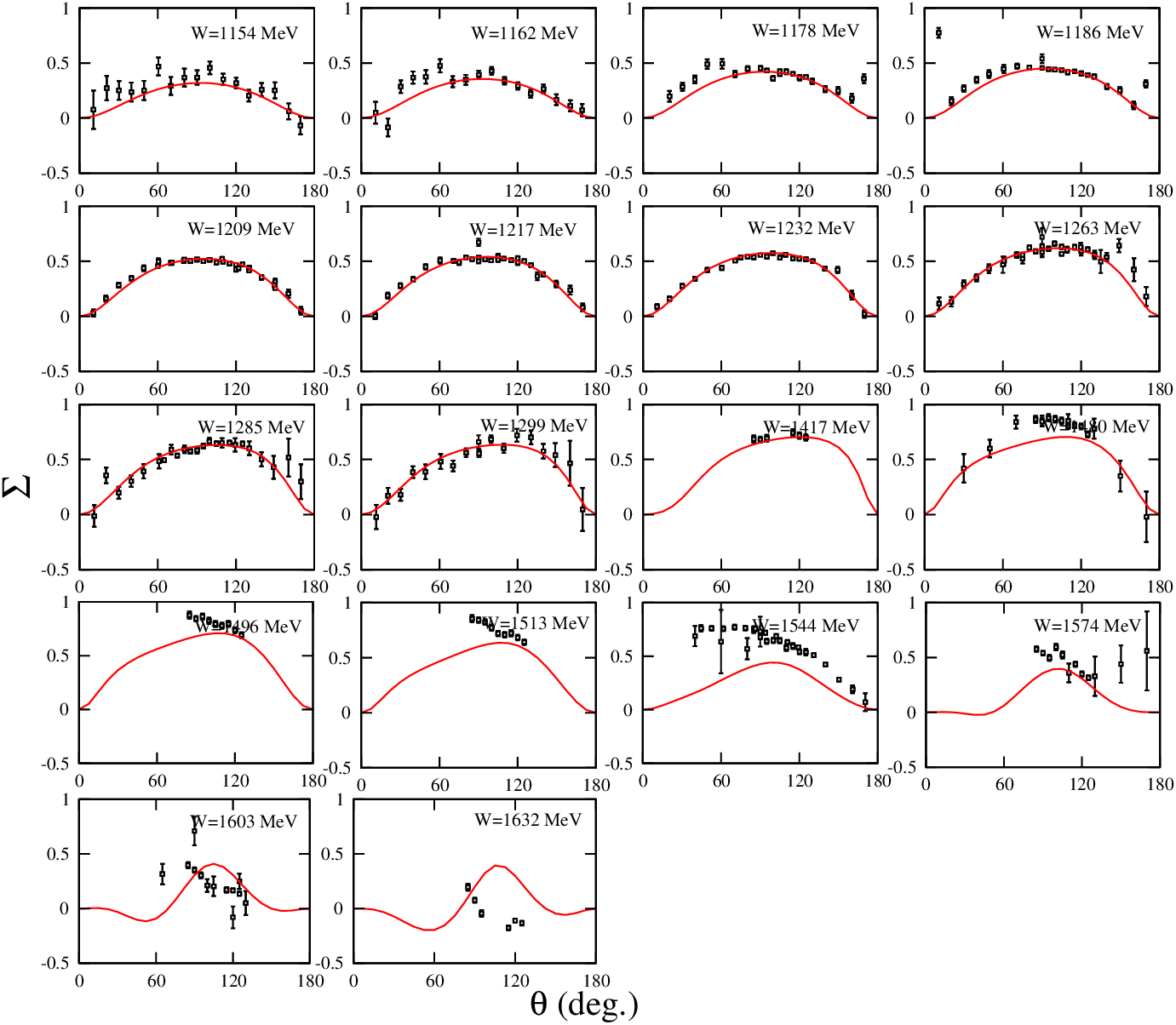}
\caption{Photon asymmetry, $\Sigma_\gamma$, for $\gamma p \to \pi^0 p$ compared 
to experimental data obtained from Ref.~\cite{saiddb}.}
\label{fig:polp0m0}
\end{figure}

\begin{figure}
\centering
\includegraphics[width=10cm,angle=-00]{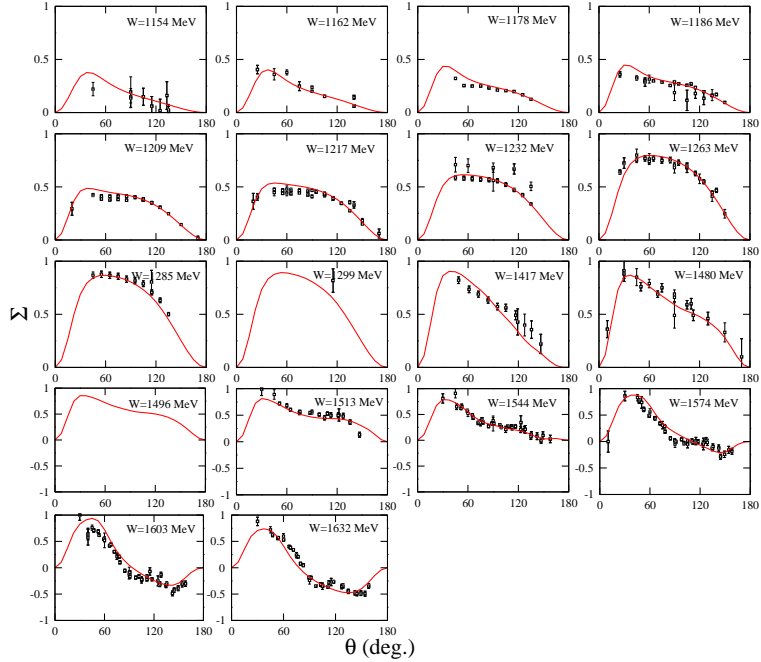}
\caption{Photon asymmetry, $\Sigma_\gamma$, for $\gamma p \to \pi^+ n$ compared 
to experimental data obtained from Ref.~\cite{saiddb}.}
\label{fig:polppm0}
\end{figure}

With the specifications given in section II, our task is to perform
$\chi^2-$fits to the available data of $\gamma N \rightarrow \pi N$
reactions up to $W = 1.65$ GeV by adjusting the bare helicity amplitudes 
$A^{J}_\lambda$ of the bare $N^*$ states included in the JLMS fit~\cite{jlms} 
to the $\pi N$ scattering data. The $\chi^2$-fits are performed by using 
MINUIT. In this first and second resonance regions, we find that
the fits to the data are mainly sensitive to the bare helicity
amplitudes listed in Table \ref{tab:para}. Other bare helicity amplitudes
are set to zero in the calculations. The quality of the resulting 
fit can be seen in Figs.~\ref{fig:dcsp0m0},~\ref{fig:dcsppm0}, 
~\ref{fig:polp0m0} and ~\ref{fig:polppm0}. The solid curves
in Figs.~\ref{fig:dcsp0m0} and~\ref{fig:dcsppm0} are the calculated
 differential cross sections ($d\sigma/d\Omega$ ) 
for $\gamma p \rightarrow \pi^0 p$ 
and $\gamma p \rightarrow \pi^+ n$, respectively, compared with the
corresponding experimental data. Similarly solid curves in 
Figs.~\ref{fig:polp0m0} and ~\ref{fig:polppm0} correspond to the obtained 
photon asymmetry ($\Sigma_\gamma$) compared to the experimental data 
for the reactions $\gamma p \rightarrow \pi^0 p$ and 
$\gamma p \rightarrow \pi^+ n$, respectively. We see that the model can 
give an overall good description of the considered data, while significant 
discrepancies with the data remain. 

We emphasize here that the determined bare helicity amplitudes listed in 
Table~\ref{tab:para} are not directly the properties associated with the 
nucleon 
resonances. They are simply the properties of the excited nucleon states 
in the $absence$ of coupling to the reaction channels. We need to identify 
the resonant positions from the partial-wave amplitudes predicted by our 
model. The dressed helicity amplitudes $\bar{A}^{J}_{\lambda}$ calculated 
at those resonance positions according to Eq.~(\ref{eq:pw-v}) can then 
be compared to the partial decay widths from the measurements, as seen in 
Eq.~(\ref{eq:width}). This is being pursued by developing~\cite{ssl}
an analytic continuation method to evaluate the reaction amplitudes in the 
complex energy plane following the dynamical coupled-channels equations 
of the MSL model~\cite{msl}.

\begin{table}[t]
\centering
\begin{tabular}{c|ccccccc} \hline\hline
   Bare  $N^*$  &  & $A_{1/2} [10^{-3}$ GeV$^{-1/2}]$ & & & $A_{3/2} [10^{-3}$ GeV$^{-1/2}]$    &  \\ \hline
$S_{11}-1$      &  & 69     & & & --       &  \\ \hline
$S_{11}-2$      &  & -17    & & & --       &  \\ \hline
$S_{31}-1$      &  & 188   & & & --       &  \\ \hline
$P_{11}-1$      &  & 23     & & & --       &  \\ \hline
$P_{13}-1$      &  & $-$64   & & &  $-$20  &  \\ \hline
$P_{33}-1$      &  & $-$78  & & & $-$132      &  \\ \hline
$D_{13}-1$      &  & 47    & & & $-$72   &  \\ \hline
$D_{15}-1$      &  & 47     & & &  32 &  \\ \hline
$D_{33}-1$      &  &  30     & & & $-$51  &  \\ \hline
$F_{15}-1$      &  &$-$97    & & & $-$63  &  \\ \hline
\hline\hline
\end{tabular}
\caption{The bare $\gamma N \rightarrow N^*$
helicity amplitudes determined from $\chi^2$-fits to the
$\gamma N \rightarrow \pi N$ data shown 
 in Figs.~\ref{fig:dcsp0m0}-\ref{fig:polppm0}.  }
\label{tab:para}
\end{table}

\begin{figure}[tb]
\vspace{20pt}
\centering
\includegraphics[width=12cm,angle=-00]{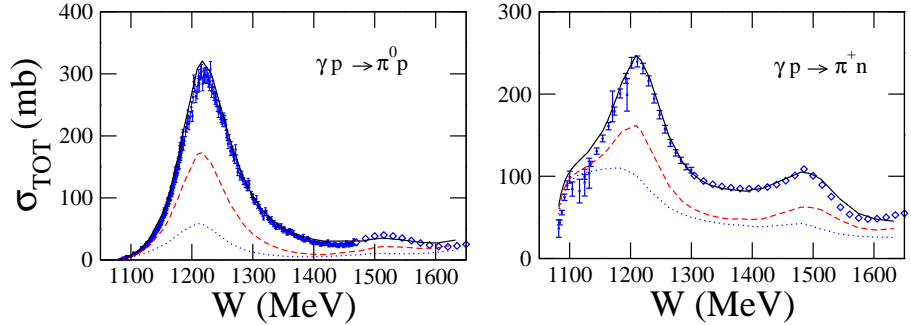}
\caption{Total cross sections. The dashed curves are obtained from
turning off all $MB$ channels except the $\pi N$
channel in the the loop integrations in the non-resonant
amplitude defined by Eq.~(\ref{eq:pw-nonr}) and the dressed
$\gamma N \rightarrow N^*$ vertex defined by Eq.~(\ref{eq:pw-v}). 
The dotted curve is obtained by neglecting the off shell effects 
in the $\pi N$ only calculation. Experimental data obtained using 
Ref.~\cite{saiddb}. The diamonds correspond to the SAID solution~\cite{said}.}
\label{fig:tcs}
\end{figure}

\begin{figure}[t]
\vspace{40pt}
\centering
\includegraphics[width=12cm,angle=-0]{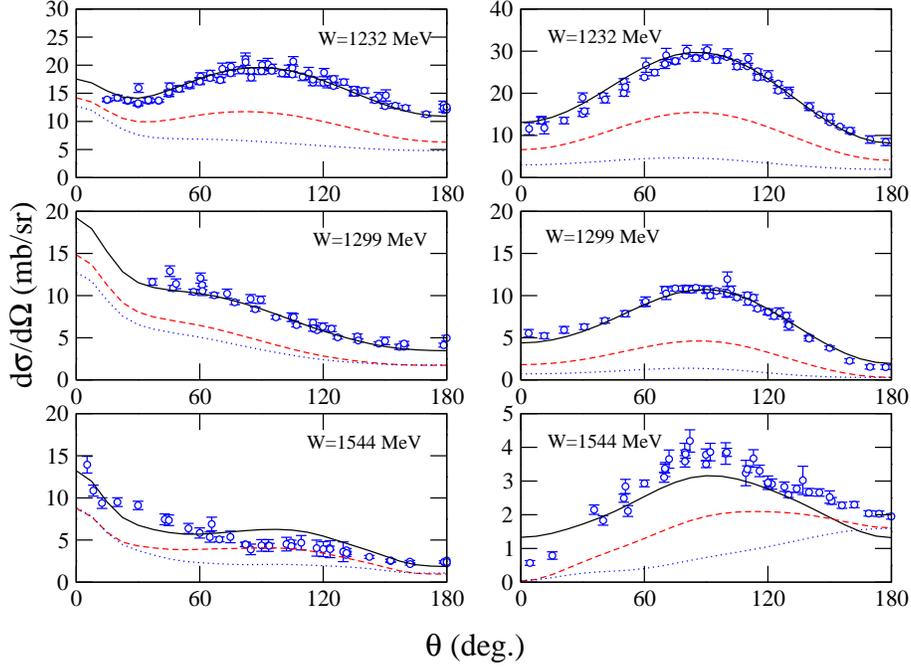}
\caption{coupled-channels effects on $d\sigma/d\Omega$
at W = 1.23, 1.30, 1.54 GeV. The dashed curves are obtained from
turning off all $MB$ channels except the $\pi N$
channel in the the loop integrations in the non-resonant
amplitude defined by Eq.~(\ref{eq:pw-nonr}) and the dressed
$\gamma N \rightarrow N^*$ vertex defined by Eq.~(\ref{eq:pw-v}). 
The dotted curve is obtained by neglecting the off shell effects 
in the $\pi N$ only calculation. Right-hand side :$\gamma p \rightarrow \pi^0 p$,
left-hand side :$\gamma p \rightarrow \pi^+ n$}
\label{fig:dcs-cc}
\end{figure}

We now turn to investigating the coupled-channels effects.
In Fig.~\ref{fig:tcs}, we see that the calculated total cross 
sections (solid curves) are in good agreement with the data.
The dashed curves are obtained when the channels $\eta N$, 
$\pi\Delta$, $\rho N$, and $\sigma N$ are turned off in the 
loop integrations in the non-resonant amplitude defined by 
Eq.~(\ref{eq:pw-nonr}) and the dressed $\gamma N \rightarrow N^*$ 
vertex defined by Eq.~(\ref{eq:pw-v}). Clearly, the coupled-channels 
effects $\gamma N \rightarrow \eta N$, $\pi\Delta$, $\rho N$, 
$\sigma N$ $\rightarrow \pi N$ can change the cross sections by 
about 10 - 20 $\%$ in the $\Delta$ (1232) region and as much as 
50 $\%$ in the $W > $1400 MeV second resonance region. The corresponding 
coupled-channels effects on the differential cross sections are 
shown in Fig.~\ref{fig:dcs-cc}. Comparing the solid and dashed 
curves, we see that the coupled-channels effects can change the 
magnitudes and shapes of $d\sigma/d\Omega$, in particular at higher 
energies.  

As discussed in the introduction, an essential feature of a 
dynamical approach is to solve integral equations 
Eqs.~(\ref{eq:teq})-(\ref{eq:veq}) which involve off-shell amplitudes. 
The off-shell effects are due to the reaction mechanisms at short 
distances which are not treated explicitly in the technically much 
simpler K-matrix coupled-channels 
models~\cite{giessen,kvi,bonn,said,maid,jlab-yeve}. If 
the off-shell effects are neglected in the $\pi N$-loop only calculations, 
we then bring the dashed curves in Figs.\ref{fig:tcs} and \ref{fig:dcs-cc} 
to the dotted curves. Clearly, the off-shell effects are very 
significant, as was also revealed in the coupled-channels 
calculations~\cite{jslt} of $KY$ photoproduction.

To disentangle the structure of nucleon resonances, it is 
important to investigate the meson cloud effect on the 
$\gamma N \rightarrow N^*$, as defined by the second term 
of Eq.~(\ref{eq:pw-v}). It can have contributions from the 
loop integrations over the $\pi N$, $\eta N$, $\pi\Delta$, 
$\rho N$, $\sigma N$ intermediate states in this calculation. 
The contribution from the $\pi N$ loop is that illustrated in 
Fig.~\ref{fig:d-vertex}. If the electromagnetic form factors associated 
with the non-resonant interaction $v_{MB,\gamma N}$ are taken from 
previous works, we can predict the $Q^2$-dependence of the meson cloud 
effect term $A^{mc,J}_{\lambda}$ of Eq.~(\ref{eq:ggn-d}). 

\begin{figure}[t]
\centering
\includegraphics[width=10cm,angle=-0]{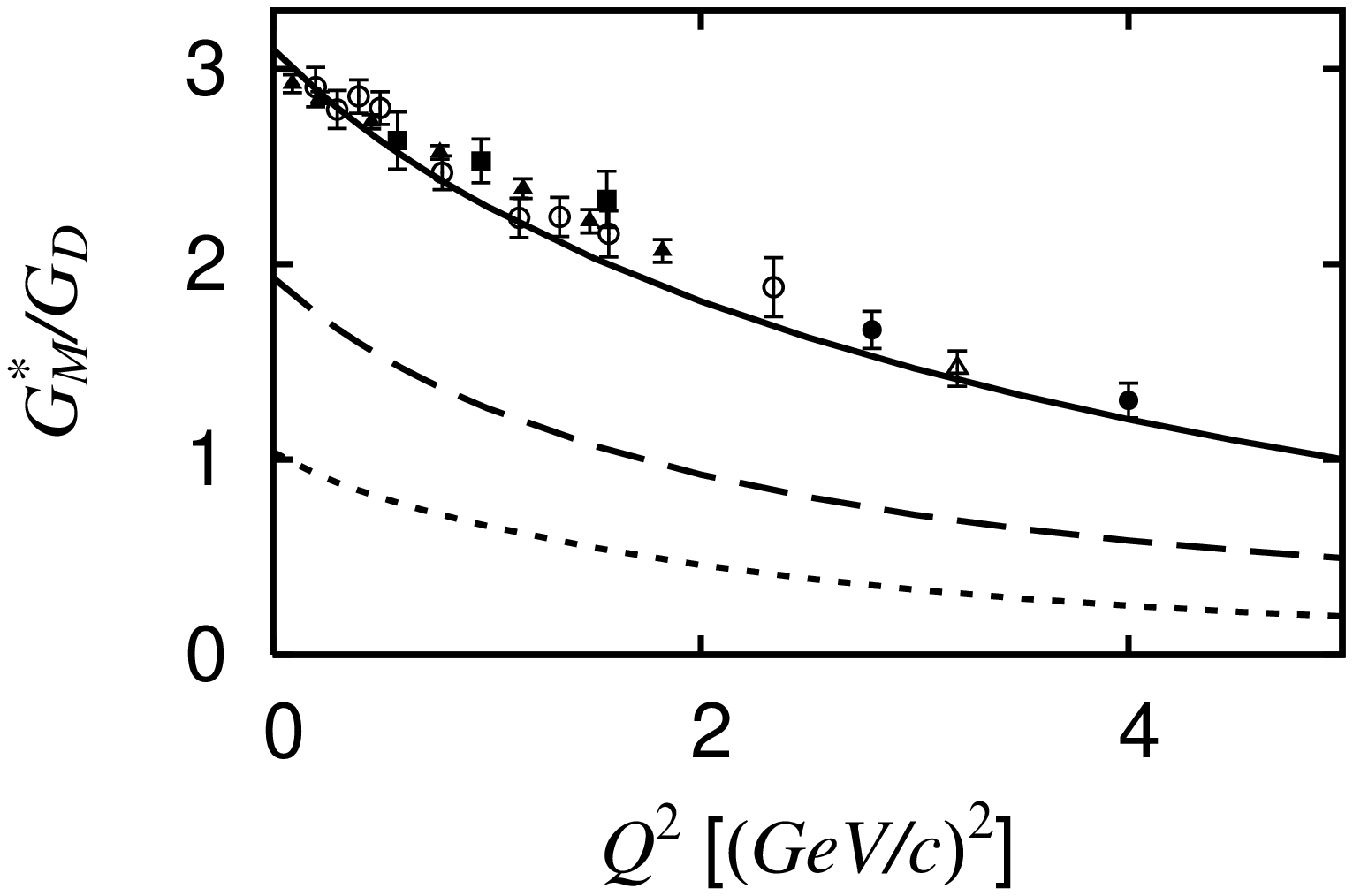}
\caption{The predicted $Q^2$-dependence of meson cloud contributions 
to the magnetic M1 form factor $G^*_M$ defined in Ref.~\cite{sl} and
$G_D (Q^2) = 1/[1+Q^2/(0.71 (GeV/c)^2)]^2$. The dashed curve in the 
lower part of the figure is the predicted full meson cloud contribution.
The dotted curve is obtained after turning off the coupled-channels 
effects due to $\pi\Delta$ and $\rho N$ channels in the loop integrations 
of Eq.~(\ref{eq:pw-v}). The data are compiled in Ref.~\cite{jlss}.\label{fig:gm}}
\end{figure}

We first investigate the $\gamma N \rightarrow \Delta$ (1232) 
transition. The resonance position of $\Delta$ (1232) is well defined
and reproduced in our calculations as can be seen in the good description 
of the cross section data near $W = 1232 $ MeV in 
Figs.~\ref{fig:dcsp0m0}-\ref{fig:tcs}. For this
isolated resonance, we can follow the procedures detailed in Ref.~\cite{sl} 
to calculate the $\gamma N \rightarrow \Delta$ (1232) magnetic form factor
$G^{*}_M (Q^2)$ from the imaginary part of the multipole amplitude
$M^{I=3/2}_{1^+}$ of $\gamma^* N \rightarrow \pi N$ reactions at $W=1232 $
MeV.  Our results are shown in
Fig.~\ref{fig:gm}. The dashed curve corresponds to the full meson cloud 
contribution from this calculation. The dotted curve is obtained by
keeping only the $\pi$-loop in the calculation. Clearly
the difference between the
dashed and dotted curves is due to the coupled-channels effects
$\gamma N \rightarrow \eta N, \pi \Delta, \rho N, \sigma N \rightarrow 
\Delta$ (1232). Here we note that the pion-loop only result (dotted curve)
is very close to the meson cloud effect predicted by the Sato-Lee 
Model\cite{sl}.  The solid curve following the data 
is just for guiding the eyes. The difference between the solid curve and the
dashed curve provide information about the $Q^2$-dependence of
the bare $\gamma N \rightarrow \Delta$ (1232) form factor which can be
used as the starting 
point of our dynamical coupled-channels analysis of pion electroproduction.

\begin{figure}[t]
\centering
\includegraphics[width=10cm,angle=-90]{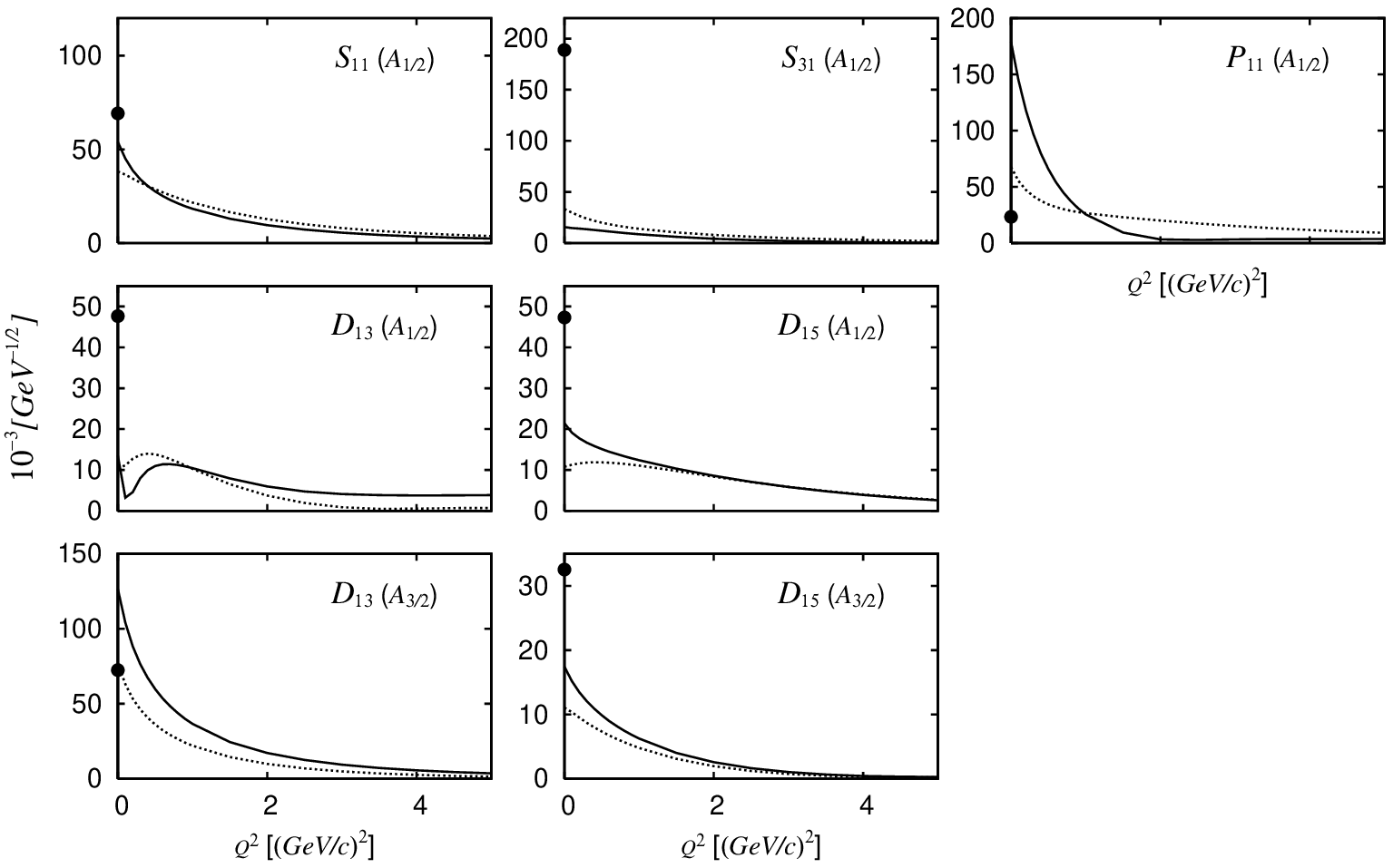}
\caption{The predicted $Q^2$-dependence of meson cloud 
contributions to the helicity amplitudes $\bar{A}^{J}_\lambda$. 
The dotted curves are from keeping only the $\pi$-loop in 
Eq.~(\ref{eq:pw-v}). The black dot corresponds to the 
absolute value of the bare helicities listed in Table.~\ref{tab:para}.}
\label{fig:cloud}
\end{figure}

For the meson cloud effects on the other $\gamma N \rightarrow N^*$ 
vertex, we calculate the second term of Eq.~(\ref{eq:pw-v}) to get 
$A^{mc,J}_{\lambda}$ by using the normalization defined by 
Eqs.~(\ref{eq:ggn})-(\ref{eq:ggn-d}). In Fig.~\ref{fig:cloud}, we show 
the predicted magnitudes of $|A^{mc,J}_{\lambda}|$ evaluated at $W = 1535$ 
MeV for $S_{11}$, $W = 1440 $ MeV for $P_{11}$, 
$W = 1520 $ MeV for $D_{13}$, $W = 1625 $ MeV for $D_{15}$, and
$W = 1620$ MeV for $S_{31}$. The solid dots at $Q^2=0$ are
the determined bare helicity amplitudes. The solid curves are from the
full calculations and the dashed curves are from only keeping the
$\pi N$ loop in Eq.~(\ref{eq:pw-v}). We see that the meson cloud 
contributions and the coupled-channels effects affect mainly the 
low $Q^2$ region.

Here we note that the results presented in Fig.~\ref{fig:cloud} 
are around the resonance positions listed by PDG, not from the
resonance pole positions which will be determined~\cite{ssl}
using an analytical continuation~\cite{ssl}. Thus the results presented 
here are only for giving some qualitative estimate of the meson 
cloud effects on $\gamma N \rightarrow N^*$ excitation. More accurate 
predictions will be published in our subsequent analysis~\cite{ssl} 
of the data of pion electroproduction.

\section{Summary and Future developments}

We have applied the dynamical coupled-channels model of 
Ref.~\cite{msl} to investigate the pion photoproduction 
reactions in the first and second nucleon resonance region. 
With the hadronic parameters of the model determined in 
a recent study~\cite{jlms} of $\pi N$ scattering data 
and the non-resonant electromagnetic couplings taken from 
the previous works, we show that the available data of 
differential cross sections and photon asymmetries of 
$\gamma N \rightarrow \pi N$ up to $W = 1.65$ GeV can be 
described to a very large extent. The only free parameters 
in the $\chi^2$-fit to the data are the bare $\gamma N\rightarrow N^*$ 
helicity amplitudes. It is found that the coupled-channels
effects can have about 10 - 20 $\%$ effects in the $\Delta$ (1232) 
resonance region, and can drastically change the magnitudes 
and shapes of the cross sections in the second resonance region. 
We also demonstrate the importance of the off-shell effects 
in a dynamical approach.
The meson cloud contributions to the $\gamma N \rightarrow N^*$ 
form factors have been predicted. For all cases, they are mainly 
in the low $Q^2$ region. For the magnetic M1 
$\gamma N \rightarrow \Delta$ (1232) form factor, the results are 
close to those of the Sato-Lee Model. The coupled-channels effects on
the meson cloud contributions are also found to be mainly in the 
low $Q^2$ region. 

The results presented here can be the starting point for analyzing 
the pion electroproduction data in the first and second resonance 
region. It will be interesting to see the bare $\gamma N \rightarrow
\Delta$ (1232) form factor which can be estimated from the differences between
the solid and dashed curves in Fig.~\ref{fig:gm} can be verified in the
analysis of the data at $W = 1232$ MeV and $Q^2 \leq 6 $ (GeV/c)$^2$. 
 For the data in the second resonance 
region, our task is to extract the bare helicity amplitudes of 
$\gamma N \rightarrow N^*$ at each $Q^2$, as was done in this work 
at the photon point $Q^2=0$. Of course the data to be fitted are much 
more extensive and we need to also determine the longitudinal components 
of the $\gamma N \rightarrow N^*$ vertex. This is being 
pursued~\cite{jlmss-1} at EBAC.

The most unsatisfactory part of this work is the uncertainties in 
determining the bare $\gamma N \rightarrow N^*$ helicity amplitudes. 
As seen in Table \ref{tab:para}, there are 16 helicity amplitudes 
adjusted in our $\chi^2$-fits to the data. Obviously, it is possible 
that there could exist other solutions to the minimization problem 
if more experimental data, in particular polarization data is included 
in the fit.
Within the MSL formulation~\cite{msl}, these bare parameters are 
related to hadron structure calculations in the absence of the couplings 
with reaction channels. A more fruitful approach is to take the 
helicity amplitudes predicted from such  hadron structure calculations 
as the starting values in our $\chi^2$-fit to avoid unphysical solutions. 
The resulting  parameters could then be used to examine that hadron 
structure calculation. However more theoretical analysis is needed to 
know precisely what kind of structure calculations are consistent with 
the MSL formulation and can be used for this purpose. Since it was 
found~\cite{jlss} that the extracted bare $\gamma N \rightarrow \Delta$ (1232)
magnetic M1 form factor is fairly consistent with the prediction of
constituent quark models, one possibility is to use the relativistic 
constituent quark model~\cite{bruno-rqm}.

To improve the agreement with the data, one necessary next step 
is to improve the hadronic parameters of the model. These parameters, 
fixed at the values from the fit to $\pi N$ scattering data, must 
be improved by performing a combined analysis of both the $\pi N$ elastic 
scattering and $\pi N \rightarrow \pi\pi N$ data. We expect that the
parameters associated with the unstable particle channels, $\pi\Delta$, 
$\rho N$, and $\sigma N$, will be refined most significantly. Thus the 
predicted coupled-channels effects on $\gamma N \rightarrow \pi N$ cross 
sections, as shown in Fig.~\ref{fig:dcs-cc} will be changed such that
the fits to the data shown in Figs.~\ref{fig:dcsp0m0}-\ref{fig:polppm0}, 
in particular in the high $W$ region, can be improved. Furthermore, this 
is also needed to extend our investigation to the third resonance region 
where the two pion production dominates and the coupled-channels effects 
through these unstable particle channels are expected to be very large. 

Finally, we would like to address the questions concerning how the 
results from JLMS and this investigation can be used to extract the 
positions and widths of nucleon resonances. The resonances positions 
are the poles of the reaction amplitudes in the complex energy plane. 
The residues of these poles can then be related to the partial decay 
widths. If these poles are identical to the zeros of the $N^*$ 
propagator $D_{i,j}(E)$ Eq.~(\ref{eq:nstar-g}), we then have the most 
desirable interpretation that the nucleon resonance is due to the coupling 
of the bare $N^*$ with the reaction channels.  The dressed vertex 
functions $\Gamma_{N^*,\gamma N}$, defined by Eq.~(\ref{eq:pw-v}), evaluated 
at these poles can be used to predict the dressed helicity amplitudes 
$\bar{A}_{\lambda}$ for calculating the partial decay widths using 
Eq.~(\ref{eq:width}). On the other hand, the poles could be from the 
non-resonant amplitude such as the term
 $t^J_{LS_N\pi N,\lambda_\gamma\lambda_N}$ of Eq.~(\ref{eq:pw-t}).
Then the identified resonances  
have nothing to do with the bare $N^*$ states and are simply due to the 
attractive meson-baryon interactions. Extraction of this resonance information 
requires developing numerical methods for solving the dynamical 
coupled-channels equations on the complex energy plane. 
Our effort in this direction will be published elsewhere~\cite{ssl}.

\begin{acknowledgments}
We would like to thank Dick Arndt for useful discussions. 
We thank Mark W. Paris for pointing out a coding error in the calculation of
the $\gamma N -> \rho N$ matrix element. B.J-D acknowledges the support of the 
Japanese Society for the Promotion 
of Science (JSPS), grant number: PE 07021. B.J-D. thanks the nuclear 
theory group at 
Osaka University for their warm hospitality. This work is supported by the 
U.S. Department of Energy, Office of Nuclear Physics Division, under 
contract No. DE-AC02-06CH11357, and Contract No. DE-AC05-060R23177 
under which Jefferson Science Associates operates Jefferson Lab,
and by the Japan Society for the Promotion of Science,
Grant-in-Aid for Scientific Research(c) 15540275. This work is also
partially supported by Grant No. FIS2005-03142 from MEC (Spain) 
and FEDER and European Hadron Physics Project RII3-CT-2004-506078. 
\end{acknowledgments}

\end{document}